\documentstyle[multicol,eqsecnum,aps]{revtex}
\renewcommand{\narrowtext}{\begin{multicols}{2}
\global\columnwidth20.5pc\noindent}
\renewcommand{\widetext}{\end{multicols}
\global\columnwidth42.5pc}
\begin{document}
\draft
\title{Ground-state properties of the two-band model for
       halogen-bridged metal complexes}
\author{Shoji Yamamoto}
\address{Department of Physics, Faculty of Science, Okayama University,
         Okayama 700-8530, Japan}
\maketitle
\begin{abstract}
   Based on a symmetry argument, we systematically reveal Hartree-Fock
broken-symmetry solutions of the one-dimensional two-band extended
Peierls-Hubbard model.
Performing numerical investigations as well, the possibility of novel
density-wave states appearing is argued.
\end{abstract}
\pacs{{\it Keywords}:
      {\it Electron-density calculations, Magnetic phase transitions}}

\narrowtext
\section{Introduction}\label{S:I}

   One-dimensional two-band models with competing electron-electron
(el-el) and electron-phonon (el-ph) interactions have been attracting
considerable interest.
Such models cover quasi-one-dimensional materials such as
halogen-bridged transition-metal (MX) linear-chain complexes and
charge-transfer (CT) salts with mixed-stacks of alternating donor
and acceptor molecules.

   One of the most interesting consequences of intrinsic multiband
effects and the competition between el-el and el-ph interactions
may be the variety of ground states
\cite{Yama29,Gamm08,Yone65,Bish51,Gamm77}.
Mixed-stack CT compounds \cite{Torr53,Iwas15,Toku05} such as
TTF-chloranil and TTeC$_1$TTF-TCNQ may be described by a half-filled
two-band model with site-off-diagonal el-ph coupling.
They are dimerized into bond-order-wave (BOW) states at low
temperatures, which is recognized as a charge-transfer-induced
spin-Peierls transition.
Ground states of MX compounds, which may be described by a
$3/4$-filled two-band model with site-diagonal el-ph coupling,
strongly depend on the constituent metal and halogen ions.
The conventional Pt compounds have the ground state
with a charge density wave (CDW) on the metal sites induced by a
dimerization of the halogen-ion sublattice, whereas recently
synthesized Ni compounds \cite{Toft61,Tori41} show a regular-chain
structure with a spin density wave (SDW) on the metal sites.
Although the interchain hydrogen-bond networks are not negligible
here, yet the competition between the CDW and SDW states may more
or less be recognized as a crossover between the Peierls and Mott
insulators.
Thus one-dimensional el-ph systems show rich phase diagrams and
may possess further ground states unrevealed.

\section{Model Hamiltonian}\label{S:H}

   We study the one-dimensional two-band extended Peierls-Hubbard
model:
\begin{eqnarray}
   {\cal H}
   &=& \sum_{n=1}^L\sum_{s=\pm}
       \bigl[\varepsilon_{\rm M}-\beta(u_n-u_{n-1})\bigr]
       a_{1:n,s}^\dagger a_{1:n,s}
       \nonumber \\
   &+& \sum_{n=1}^L\sum_{s=\pm}
       \varepsilon_{\rm X} a_{2:n,s}^\dagger a_{2:n,s}
       \nonumber \\
   &-& \sum_{n=1}^L\sum_{s=\pm}
       \bigl[(t-\alpha u_n)a_{1:n,s}^\dagger a_{2:n,s}
       \nonumber \\
   &+& (t+\alpha u_{n-1})a_{1:n,s}^\dagger a_{2:n-1,s}
       +{\rm H.c.}\bigr]
       \nonumber \\
   &+& U_{\rm M}\sum_{n=1}^L
       a_{1:n,+}^\dagger a_{1:n,+}
       a_{1:n,-}^\dagger a_{1:n,-}
       \nonumber \\
   &+& U_{\rm X}\sum_{n=1}^L
       a_{2:n,+}^\dagger a_{2:n,+}
       a_{2:n,-}^\dagger a_{2:n,-}
       \nonumber \\
   &+& V\sum_{n=1}^L\sum_{s,s'=\pm}
       \bigl[
        a_{1:n,s}^\dagger a_{1:n,s} a_{2:n,s'}^\dagger a_{2:n,s'}
       \nonumber \\
   &+& a_{1:n,s}^\dagger a_{1:n,s} a_{2:n-1,s'}^\dagger a_{2:n-1,s'}
       \bigr]
    +  \frac{K}{2}\sum_{n=1}^L
       u_n^2 \,.
   \label{E:Hn}
\end{eqnarray}
Here we assume the Hamiltonian to describe MX chains.
Now the notation of the Hamiltonian (\ref{E:Hn}) is recognized as
follows:
$a_{1:n,s}^\dagger$ and $a_{2:n,s}^\dagger$ are the creation
operators of an electron with spin $s$ in the metal $d_{z^2}$ and
halogen $p_z$ orbitals at the unit cell $n$, respectively, $u_n$ the
chain-direction displacement from the uniform lattice spacing of the
halogen atom at the unit cell $n$, and $L$ the number of unit cells;
$\varepsilon_{\rm M}$ and $\varepsilon_{\rm X}$
are the energies of the metal $d_{z^2}$ and halogen $p_z$
orbitals on isolated atoms, respectively,
$t$ the transfer energy of hopping between these levels,
$U_{\rm M}$, $U_{\rm X}$, and $V$
the el-el interactions on the metal atoms, on the halogen atoms, and
between the neighboring metal and halogen atoms, respectively;
$\alpha$, $\beta$, and $K$ denote
the site-off-diagonal and site-diagonal el-ph coupling constants
and the elastic constant, respectively.
We note that the model is rather general and may essentially be
applied to various materials.
For example, replacing the $d_{z^2}$ and $p_z$ orbitals by $p\pi$
orbitals of donor and acceptor molecules, respectively, we can
immediately switch our argument to organic mixed-stack CT compounds.
We compactly rewrite Hamiltonian (\ref{E:Hn}) as
\begin{eqnarray}
   {\cal H}
   &=& \sum_{i,j}\sum_{k,q}\sum_{s,s'}
       \langle i:k+q,s\vert t\vert j:k,s'\rangle
       a_{i:k+q,s}^\dagger a_{j:k,s'}
       \nonumber \\
   &+& \frac{1}{2}\sum_{i,j,m,n}\sum_{k,k',q}\sum_{s,s',t,t'}
       \nonumber \\
   & & \langle i:k+q,s;m:k',t\vert v\vert j:k,s';n:k'+q,t'\rangle
       \nonumber \\
   &\times&
       a_{i:k+q,s}^\dagger a_{m:k',t}^\dagger
       a_{n:k'+q,t'} a_{j:k,s'}
    +  \frac{K}{2}
       \sum_k u_k u_k^* \,,
   \label{E:Hk}
\end{eqnarray}
where
\begin{eqnarray}
   && \langle i:k+q,s\vert t\vert j:k,s'\rangle
     =\langle i:k+q\vert t\vert j:k\rangle\delta_{ss'} \,,
      \label{E:intt}
      \nonumber \\
   && \langle i:k+q,s;m:k',t\vert v\vert j:k,s';n:k'+q,t'\rangle
      \nonumber \\
   && \quad
     =\langle i:k+q;m:k'\vert v\vert j:k;n:k'+q\rangle
      \delta_{ss'}\delta_{tt'} \,,
      \label{E:intv}
\end{eqnarray}
are specified as
\begin{eqnarray}
   && \langle 1:k+q\vert t\vert 1:k\rangle
     ={\widetilde\varepsilon}_{\rm M}\delta_{q0}
     -\frac{2{\rm i}\beta}{\sqrt{L}}u_q^*
      \sin\left(\frac{q}{2}\right),
      \nonumber \\
   && \langle 2:k+q\vert t\vert 2:k\rangle
     ={\widetilde\varepsilon}_{\rm X}\delta_{q0} \,,
      \nonumber \\
   && \langle 1:k+q\vert t\vert 2:k\rangle
     =\frac{2{\rm i}\alpha}{\sqrt{L}}u_q^*
      \sin\biggl(\frac{k+q}{2}\biggr)
     -2t\delta_{q0}
      \cos\biggl(\frac{k}{2}\biggr),
      \nonumber \\
   && \langle 1:k+q;1:k'\vert v\vert 1:k;1:k'+q\rangle
     =\frac{U_{\rm M}}{L} \,,
      \nonumber \\
   && \langle 2:k+q;2:k'\vert v\vert 2:k;2:k'+q\rangle
     =\frac{U_{\rm X}}{L} \,,
      \nonumber \\
   && \langle 1:k+q;2:k'\vert v\vert 1:k;2:k'+q\rangle
     =\frac{2V}{L}\cos\left(\frac{q}{2}\right).
   \label{E:int}
\end{eqnarray}
with the renormalized on-site affinities
${\widetilde\varepsilon}_{\rm M}=\varepsilon_{\rm M}-U_{\rm M}/2$
and
${\widetilde\varepsilon}_{\rm X}=\varepsilon_{\rm X}-U_{\rm X}/2$.

   The symmetry group of the system is generally represented as
$
   {\bf G}={\bf P}\times{\bf S}\times{\bf T}
$,
where ${\bf P}={\bf L}_1\land{\bf C}_2$ is the space group of a linear
chain with the one-dimensional translation group ${\bf L}_1$ whose
basis vector is the unit-cell translation,
${\bf S}$ the group of spin-rotation, and
${\bf T}$ the group of time reversal.
Group actions on the creation operators are readily found in
Ref. \cite{Ozak55}.
Let $\check{G}$ denote the irreducible representations of {\bf G}
over the real number field, where their representation space is spanned
by the Hermitian operators \{$a_{i:k,s}^\dagger a_{j:k',s'}$\}.
There is a one-to-one correspondence \cite{Ozak14} between
$\check{G}$ and broken-symmetry phases with a definite ordering vector.
Any representation $\check{G}$ is obtained as a Kronecker product of
the irreducible representations of ${\bf P}$, ${\bf S}$, and ${\bf T}$
over the real number field:
$
   \check{G}=\check{P}\otimes\check{S}\otimes\check{T} \,.
$
$\check{P}$ is characterized by an ordering vector $q$ in the
Brillouin zone and an irreducible representation of its little group
${\bf P}(q)$, and is therefore labeled $q\check{P}(q)$.
The relevant representations of ${\bf S}$ are
$
   \check{S}^0(u(\mbox{\boldmath$e$},\theta))
      =1
$ and
$
   \check{S}^1(u(\mbox{\boldmath$e$},\theta))
      =O(u(\mbox{\boldmath$e$},\theta))
$, whereas ones of ${\bf T}$ are
$
   \check{T}^0(t)=1
$
$
   \check{T}^1(t)=-1
$, where
$
   u(\mbox{\boldmath$e$},\theta)
   =\sigma^0\cos(\theta/2)
   -(\mbox{\boldmath$\sigma$}\cdot\mbox{\boldmath$e$})
    \sin(\theta/2)
$
represents the spin rotation of angle $\theta$ around an axis
$\mbox{\boldmath$e$}$ and
$O(u(\mbox{\boldmath$e$},\theta))$ is the $3\times 3$
orthogonal matrix satisfying
$
   u(\mbox{\boldmath$e$},\theta)
   \mbox{\boldmath$\sigma$}^\lambda
   u^\dagger(\mbox{\boldmath$e$},\theta)
      =\sum_{\mu=x,y,z}
       [O(u(\mbox{\boldmath$e$},\theta))]_{\lambda\mu}
       \mbox{\boldmath$\sigma$}^\mu \ \
       (\lambda=x,\,y,\,z)
$,
with the $2\times 2$ unit matrix $\sigma^0$ and
the Pauli-matrices vector
$\mbox{\boldmath$\sigma$}=(\sigma^x,\sigma^y,\sigma^z)$.
The representations
$\check{P}\otimes\check{S}^0\otimes\check{T}^0$,
$\check{P}\otimes\check{S}^1\otimes\check{T}^1$,
$\check{P}\otimes\check{S}^0\otimes\check{T}^1$, and 
$\check{P}\otimes\check{S}^1\otimes\check{T}^0$
correspond to charge-density-wave (CDW), spin-density-wave (SDW),
charge-current-wave (CCW), and spin-current-wave (SCW) states,
respectively.
We leave out current-wave states in our argument, because here in
one dimension all of them but one-way uniform-current states break
the charge- or spin-conservation law.
We consider two ordering vectors $q=0$ and $q=\pi$, which are labeled
$\mit\Gamma$ and $X$, respectively.
Thus we treat the instabilities characterized as
${\mit\Gamma}\check{P}({\mit\Gamma})
 \otimes\check{S}^0\otimes\check{T}^0$,
$X\check{P}(X)
 \otimes\check{S}^0\otimes\check{T}^0$,
${\mit\Gamma}\check{P}({\mit\Gamma})
 \otimes\check{S}^1\otimes\check{T}^1$, and
$X\check{P}(X)
 \otimes\check{S}^1\otimes\check{T}^1$.
Here
$\check{P}({\mit\Gamma})$ and $\check{P}(X)$ are either $A$
($C_2$-symmetric) or $B$ ($C_2$-antisymmetric) representation of
${\bf C}_2$ because ${\bf P}({\mit\Gamma})={\bf P}(X)={\bf C}_2$
in the present system.

\section{Broken Symmetry Solutions}\label{S:BSS}

   In the HF approximation, the Hamiltonian (\ref{E:Hk}) is replaced
by
\begin{eqnarray}
   {\cal H}_{\rm HF}
     &=&\sum_{i,j}\sum_{k,s,s'}\sum_{\lambda=0,z}
       \bigl[
         x_{ij}^{\lambda}({\mit\Gamma};k)
         a_{i:k,s}^\dagger a_{j:k,s'}
     \\ \nonumber
     &+&x_{ij}^{\lambda}(X;k)
         a_{i:k+\pi,s}^\dagger a_{j:k,s'}
       \bigr]
       \sigma_{ss'}^\lambda\,,
   \label{E:HHF}
\end{eqnarray}
with
\begin{eqnarray}
   &&
   x_{ij}^0({\mit\Gamma};k)
     =\langle i:k\vert t\vert j:k\rangle
     +\sum_{m,n}\sum_{k'}
      \rho_{nm}^0({\mit\Gamma};k')
   \nonumber \\
   &&
   \qquad\times
   \bigl[
    2\langle i:k;m:k'\vert v\vert j:k;n:k'\rangle
   \nonumber \\
   &&
   \qquad\quad
    -\langle i:k;m:k'\vert v\vert n:k';j:k\rangle
   \bigr]\,,
   \nonumber \\
   &&
   x_{ij}^z({\mit\Gamma};k)
     =-\sum_{m,n}\sum_{k'}
      \rho_{nm}^z({\mit\Gamma};k')
   \nonumber \\
   &&
   \qquad\times
      \langle i:k;m:k'\vert v\vert n:k';j:k\rangle
   \nonumber \\
   &&
   x_{ij}^0(X;k)
     =\langle i:k+\pi\vert t\vert j:k\rangle
     +\sum_{m,n}\sum_{k'}
      (-1)^{\delta_{m2}}
   \nonumber \\
   &&
   \qquad\times
   \rho_{nm}^0({\mit\Gamma};k')
   \bigl[
    2\langle i:k+\pi;m:k'\vert v\vert j:k;n:k'+\pi\rangle
   \nonumber \\
   &&
   \qquad\qquad\qquad\quad
    -\langle i:k+\pi;m:k'\vert v\vert n:k'+\pi;j:k\rangle
   \bigr]\,,
   \nonumber \\
   &&
   x_{ij}^z(X;k)
   =-\sum_{m,n}\sum_{k'}
          (-1)^{\delta_{m2}}
          \rho_{nm}^z(X;k')
   \nonumber \\
   &&
   \qquad\times
          \langle i:k+\pi;m:k'\vert v\vert n:k'+\pi;j:k\rangle\,.
   \label{E:SCF}
\end{eqnarray}
Here we have introduced the density matrices
\begin{eqnarray}
   \rho_{ij}^\lambda({\mit\Gamma};k)
      &=& \frac{1}{2}\sum_{s,s'}
          \langle a_{j:k,s}^\dagger a_{i:k,s'}\rangle_{\rm HF}
          \sigma_{ss'}^\lambda \,,
   \nonumber \\
   \rho_{ij}^\lambda(X;k)
      &=& \frac{1}{2}\sum_{s,s'}
          \langle a_{j:k+\pi,s}^\dagger a_{i:k,s'}\rangle_{\rm HF}
          \sigma_{ss'}^\lambda \,,
   \label{E:rho}
\end{eqnarray}
where $\langle\cdots\rangle_{\rm HF}$ denotes the quantum average in
a HF eigenstate.
${\cal H}_{\rm HF}$ can be decomposed into spatial-symmetry-definite
components \cite{Yama43}.
Once a broken-symmetry Hamiltonian is given, its invariance group
is defined.
Keeping in mind that the density matrices characteristic of the
given Hamiltonian should also be invariant for its invariance group,
we can completely determine qualitative properties of the solution
\cite{Ozak55}.

   We are interested in 
the local charge density on the site $i$ at the $n$th unit cell,
$
   d_{i:n}
     \equiv
     \sum_s\langle a_{i:n,s}^\dagger a_{i:n,s}\rangle_{\rm HF}
$,
the local spin density on the site $i$ at the $n$th unit cell,
$
   s_{i:n}^z
     \equiv
     \frac{1}{2}
     \sum_{s,s'}
     \langle a_{i:n,s}^\dagger a_{i:n,s'}\rangle_{\rm HF}
     \sigma_{ss'}^z
$,
the complex bond order between the site $i$ at the $n$th unit cell
and the site $j$ at the $m$th unit cell,
$
   p_{i:n;j:m}
     \equiv
     \sum_s
     \langle a_{i:n,s}^\dagger a_{j:m,s}\rangle_{\rm HF}\,,
$,
and the complex spin bond order between the site $i$ at the $n$th
unit cell and the site $j$ at the $m$th unit cell,
$
   t_{i:n;j:m}^z
     \equiv
     \frac{1}{2}
     \sum_{s,s'}
     \langle a_{i:n,s}^\dagger a_{j:n,s'}\rangle_{\rm HF}
     \sigma_{ss'}^z
$.
The halogen-atom displacements $u_n$ are self-consistently determined
so as to minimize the HF energy 
$E_{\rm HF}\equiv\langle{\cal H}\rangle_{\rm HF}$.
All the solutions are explicitly written down elsewhere \cite{Yama43},
while we schematically show them in Fig. \ref{F:BSP}.

\section{Numerical Calculation and Discussion}\label{S:NCD}

   We show in Fig. \ref{F:PhD1} phase diagrams at $3/4$ band
filling, where $q=\pi$ states are predominantly stabilized.
With the increase of $\varepsilon_0$, density waves on the halogen
sites are generally reduced and single-band models come to be
justified.
The phase boundary between M-SDW and X-SDW is roughly given by
$U_{\rm X}-U_{\rm M}=1.8\varepsilon_0$, where
$\varepsilon_0\equiv\varepsilon_{\rm M}-\varepsilon_{\rm X}$.
In order to realize X-SDW ground states, a small enough
$\varepsilon_0$ and a large enough $U_{\rm X}$ are necessary.
We show in Fig. \ref{F:PhD2} phase diagrams at half band
filling, where $q=0$ states are relatively stabilized.
The off-site el-ph coupling $\alpha$ stabilizes BOW, whereas
the off-site Coulomb interaction $V$ is unfavorable to that.

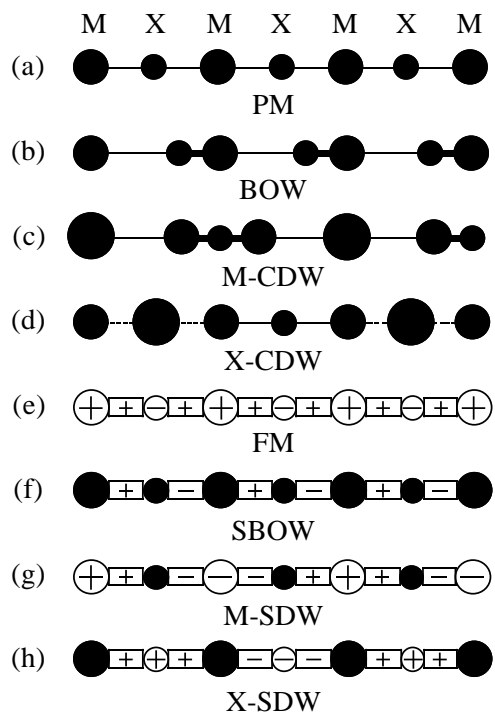
\begin{figure}
\caption{Schematic representation of possible density-wave states,
         where the variety of circles and segments qualitatively
         represents the variation of local charge densities and
         bond orders, respectively, whereas the signs $\pm$ in
         circles and strips describe the alternation of local spin
         densities and spin bond orders, respectively.
         Circles shifted from their equidistant location
         qualitatively represent X-atom displacements.
         Identifying the present system, for example, with
         halogen-bridged metal complexes, each phase is characterized
         as follows:
         (a) Paramagnetism;
         (b) Electron-phonon bond order wave;
         (c) Metal charge density wave accompanied by an
             electron-phonon bond order wave;
         (d) Halogen charge density wave accompanied by a purely
             electronic bond order wave;
         (e) Ferromagnetism with uniform spin bond orders;
         (f) Spin bond order wave;
         (g) Metal spin density wave accompanied by a
             spin bond order wave;
         (h) Halogen spin density wave accompanied by a
             spin bond order wave.}
\label{F:BSP}
\end{figure}

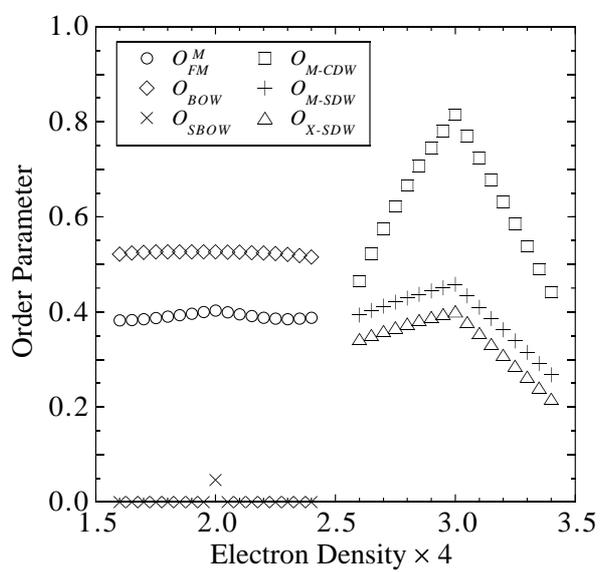
\begin{figure}
\caption{Order parameters as functions of band filling.}
\label{F:OP}
\end{figure}

   Let us observe the doping dependence of the states.
In Fig. \ref{F:OP} we plot macroscopic order parameters,
$
   O_{\rm M\mbox{-}CDW}
     =\frac{1}{L}
      \sum_n
      (-1)^n d_{1:n}
$,
$
   O_{\rm M\mbox{-}SDW}
     =\frac{1}{L}
      \sum_n
      (-1)^n s_{1:n}^z
$,
$
   O_{\rm X\mbox{-}SDW}
     =\frac{1}{L}
      \sum_n
      (-1)^n s_{2:n}^z
$,
$
   O_{\rm BOW}
     =\frac{1}{2L}
      \sum_n
      \left(
       p_{1:n;2:n-1}-p_{1:n;2:n}
      \right)
$,
$
   O_{\rm SBOW}
     =\frac{1}{2L}
      \sum_n
      \left(
       t_{1:n;2:n-1}-t_{1:n;2:n}
      \right)
$,
$
   O_{\rm FM}^{\rm M}
     =\frac{1}{L}
      \sum_n
      s_{1:n}^z
$, and
$
   O_{\rm FM}^{\rm X}
     =\frac{1}{L}
      \sum_n
      s_{2:n}^z
$
as functions of band filling, where the parametrization is as
follows:
$U_{\rm D}/t=4.0$, $U_{\rm X}/t=2.0$ for M-CDW,
$U_{\rm D}/t=8.0$, $U_{\rm X}/t=3.0$ for M-SDW, and
$U_{\rm D}/t=3.0$, $U_{\rm X}/t=8.0$ for X-SDW
with common values
$\varepsilon_0/t=1.0$,
$V/t=1.0$,
$\alpha/(Kt)^{1/2}=0.1$, and
$\beta/(Kt)^{1/2}=1.0$;
$U_{\rm D}/t=6.0$, $U_{\rm X}/t=4.0$, $V/t=1.0$ for BOW,
$U_{\rm D}/t=10.0$, $U_{\rm X}/t=9.0$, $V/t=2.0$ for SBOW, and
$U_{\rm D}/t=8.0$, $U_{\rm X}/t=3.0$, $V/t=1.0$ for FM
with common values
$\varepsilon_0/t=1.0$,
$\alpha/(Kt)^{1/2}=0.8$, and
$\beta/(Kt)^{1/2}=0.1$.
Now we explicitly find that $X$-phases are most
stabilized at $3/4$ band filling, while ${\mit\Gamma}$ ones arround
half band filling.
Though SBOW does not appear in the ground-state phase diagrams shown
here, it seems to exist in the vicinity of half band filling with
very strong on-site Coulomb repulsions.
Even if the solutions are scarcely stabilized into a ground state,
they can still be relevant in the ground-state correlations.
We note, for example, that low-lying solitonic excitations in a SDW
ground state induce SBOW domains around their centers \cite{Shim07}.
We note in addition that $O_{\rm FM}^{\rm M}$ and
$O_{\rm FM}^{\rm X}$ have opposite signs, which means spins on the
metal and halogen sites are antiparallel to each other
(Fig.$\!$ \ref{F:BSP}e).

   We finally stress the wide applicability of the present approach.
Our approach never fails to reveal all the possible broken-symmetry
phases.
We hope that the present argument will motivate further chemical,
as well as theoretical, explorations in low-dimensional el-ph
systems.

\begin{figure}
\caption{Ground-state phase diagrams at $3/4$ band filling:
         (a) $\varepsilon_0/t_0=1.0$ and (b) $\varepsilon_0/t_0=2.0$
         with common values,
         $V/t=1.0$,
         $\alpha/(Kt)^{1/2}=0.1$, and
         $\beta/(Kt)^{1/2}=1.0$.}
\label{F:PhD1}
\end{figure}

\begin{figure}
\caption{Ground-state phase diagrams at half band filling:
         (a) $V/t=1.0$ and (b) $V/t=2.0$ with common values,
         $\varepsilon_0/t_0=1.0$,
         $\alpha/(Kt)^{1/2}=0.5$, and
         $\beta/(Kt)^{1/2}=0.1$.}
\label{F:PhD2}
\end{figure}

\widetext

\begin{references}

\bibitem{Yama29}
   S. Yamamoto {\it et al.},
      Solid State Commun. {\bf 83} (1992) 329;
                          {\bf 83} (1992) 335.

\bibitem{Gamm08}
   J. T. Gammel {\it et al.},
      Phys. Rev. B {\bf 45} (1992) 6408;
   S. W. Weber-Milbrodt {\it et al.},
      {\it ibid.} {\bf 45} (1992) 6435.

\bibitem{Yone65}
   K. Yonemitsu{\it et al.},
      Phys. Rev. B {\bf 47} (1993) 8065.

\bibitem{Bish51}
   A. R. Bishop {\it et al.},
      Synth. Met. {\bf 29} (1989) F151;
   I. Batisti\'c {\it et al.},
      Phys. Rev. B {\bf 44} (1991) 13228.

\bibitem{Gamm77}
   J. T. Gammel {\it et al.},
      Synth. Met. {\bf 55}-{\bf 57} (1993) 3377;
   H. R\"oder {\it et al.},
      Phys. Rev. Lett. {\bf 70} (1993) 3498.

\bibitem{Torr53}
   J. B. Torrance {\it et al.},
      Phys. Rev. Lett. {\bf 46} (1981) 253;
   J. B. Torrance {\it et al.},
      Phys. Rev. Lett. {\bf 47} (1981) 1747.

\bibitem{Iwas15}
   N. Iwasawa {\it et al.},
      Chem. Lett. (1988) 215.

\bibitem{Toku05}
   Y. Tokura {\it et al.},
      Phys. Rev. Lett. {\bf 63} (1989) 2405.

\bibitem{Toft61}
   H. Toftlund {\it et al.},
      Inorg. Chem. {\bf 23} (1984) 4261.

\bibitem{Tori41}
   K. Toriumi {\it et al.},
      J. Am. Chem. Soc. {\bf 111} (1989) 2341.

\bibitem{Ozak55}
   M. Ozaki,
      Int. J. Quantum Chem. {\bf 42} (1992) 55;
   S. Yamamoto {\it et al.},
      {\it ibid.} {\bf 44} (1992) 949.

\bibitem{Ozak14}
   M. Ozaki,
      J. Math. Phys. {\bf 26} (1985) 1514.

\bibitem{Yama43}
   S. Yamamoto,
      Phys. Lett. A (cond-mat/9806343).

\bibitem{Shim07}
   Y. Shimoi {\it et al.},
      Solid State Commun. {\bf 82} (1992) 407.

\end{references}
\end{document}